# Ultrasound modulated optical tomography in scattering media: flux filtering based on persistent spectral hole burning in the optical diagnosis window

Caroline Venet[1,2,*], Maïmouna Bocoum[1], Jean-Baptiste Laudereau[1], Thierry Chaneliere[2], François Ramaz[1], Anne Louchet-Chauvet[2]

[1]Institut Langevin, Ondes et Images, ESPCI ParisTech, PSL Research University, CNRS UMR 7587, INSERM U979, Université Paris VI Pierre et Marie Curie, 1 rue Jussieu, 75005 Paris, France
[2] Laboratoire Aimé Cotton, CNRS, Univ. Paris-Sud, ENS Cachan, Université Paris-Saclay, Bât.505, Campus d'Orsay, 91400 Orsay France
*Corresponding author: caroline.venet@espci.fr

**Ultrasound modulated optical tomography (UOT) is a powerful imaging technique to discriminate healthy from unhealthy biological tissues based on their optical signature. Among the numerous detection techniques developed for acousto-optic imaging, only those based on spectral filtering are intrinsically immune to speckle decorrelation. This paper reports on UOT imaging based on spectral hole burning in Tm:YAG crystal under a moderate magnetic field (200G) with a well-defined orientation. The deep and long-lasting holes translate into a more efficient UOT imaging with a higher contrast and faster imaging frame rate. We demonstrate the potential of this method by imaging calibrated phantom scattering gels.**

http://dx.doi.org/10.1364/OL.43.003993

Non-invasive optical imaging is an active field of research because local absorption and scattering of tissues are critical to medical diagnosis. For example, histology of malignant breast tumors shows they have an optical signature which differs from benign cysts [1]. Optical imaging of biological tissues at depths greater than a few mm is however challenging because of light multiple scattering, making imaging techniques relying on ballistic light inapplicable. Over the last twenty years, ultrasound modulated optical tomography (UOT) has emerged as one of the bimodal-imaging technique allowing to bypass this limitation. UOT exploits the acousto-optic effect [2] which occurs between ballistic ultrasounds (US) focused along a chosen direction and diffused light. As it propagates, an US burst of carrier frequency $f_{US}$ on the MHz range modulates both the refractive index and the scattering particle positions [2]. This results in the creation of ultrasonically tagged photons shifted from the carrier frequency $f_L$ by $\pm f_{US}$ [3] [see Figure 1(a)]. By filtering these tagged photons, we retrieve a signal proportional to the local optical irradiance along the US path, with the spatial resolution of the US [4]. The ratio of tagged photons to the total number of photons lies between $10^{-3}$ and $10^{-4}$ with short bursts [5].

Two parameters are critical for imaging applications. First, because of scatterers random motion, the technique used to detect the tagged photons must handle speckle decorrelation. Second, the signal to noise ratio (SNR) should be maximal to ensure a large imaging frame rate (>Hz). A good SNR is reached when the tagged and transmitted photons are effectively discriminated for large detection *etendue*.

The most mature techniques for the detection of tagged photons are based on self-adaptive wavefront holography. Holographic response times close to 1ms or below have been reported, which is promising but still insufficient for *in vivo* imaging [6-7]. Digital holography is far less sensitive to speckle decorrelation [8]. However, the CCD camera needs to process multiple speckles in parallel thus imposes a frame rate too slow for real-time tracking of the US propagation. As opposed to holographic techniques, the confocal Fabry-Perot detection consists in spectrally filtering the tagged photons. It is intrinsically immune to speckle decorrelation but operates with limited *etendue* [9].

Spectral Hole Burning (SHB) allows, among other applications, the generation of narrow-band filters, which can discriminate ultrasonically tagged photons over a large *etendue*. In rare-earth doped crystals at cryogenic temperature, the inhomogeneous absorption linewidth is at least one order of magnitude larger that the homogeneous linewidth. This is why narrow spectral classes of atoms within the inhomogeneous broadening can be selected with SHB. To burn a spectral hole, a spectrally narrow laser saturates the homogeneous line absorption thereby creating a transparency window in the absorption profile, called spectral hole. Biological imaging requires the SHB transition to be in the optical diagnosis window [10]. Among many of the materials exhibiting the SHB property, $Tm^{3+}$ and $Nd^{3+}$ doped crystals operate at appropriate wavelengths (793nm and 883nm respectively).

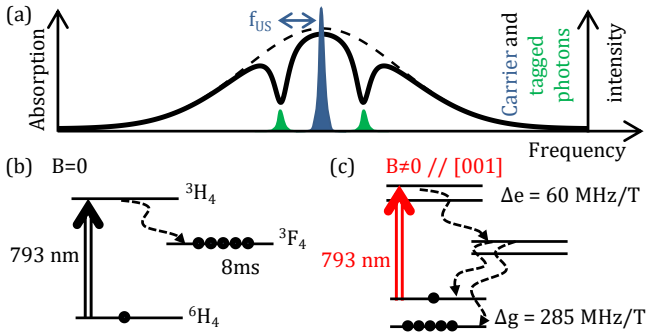

Fig. 1. (a) Spectral filtering of UOT tagged photons. Dashed line: initial absorption spectrum. Solid line: absorption spectrum after spectral holeburning at two frequencies. Colored area: spectrum of the light coming out of the scattering medium (blue: photons at the carrier frequency and green: tagged photons). $f_{US}$ is the ultrasound frequency (typically 5 MHz). (b) Tm:YAG level scheme in the absence of magnetic field. Optically excited atoms are stored in the metastable $^3F_4$ level for 8ms. (c) Level scheme under magnetic field: optically excited atoms are stored in one of the two ground state sublevels with a 30s lifetime. Δg and Δe are the Zeeman splitting dependence of the ground and excited levels, respectively [19].

UOT-SHB in Tm-doped YAG has been reported in [11-13]. Under laser excitation the resonant atoms at 793nm in Tm:YAG are promoted to the $^3H_4$ excited state, as depicted in Figure 1(b) without magnetic field. In about 200μs [14], 30% of the atoms relax straight to the ground state while 70% remain in the metastable level $^3F_4$, whose lifetime is 8ms [14]. This time defines the spectral hole lifetime, which is a key parameter for UOT because it will set the maximum number of signals acquired for a given US burst repetition rate. Because refreshing the hole is a time consuming step, a longer lifetime would in overall speed up the imaging process.

If a ground state level substructure exists (hyperfine or nuclear Zeeman structure), the hole lifetime can be extended much above the excited state lifetime, typically above a few seconds. This is called the Persistent spectral hole burning (PSHB) regime [15], and allows a more efficient optical pumping process, together with deeper holes. This was demonstrated for example in praseodymium-doped $Y_2SiO_5$, where a hole lifetime of several thousand seconds was observed at 606nm [16]. Additionally, the optical dispersion due to the steep absorption variation induces a slowing of the tagged photons going through the filter compared to untagged photons. This so-called slow light approach has been used to enhance the filter sensitivity. So far, most significant PSHB-UOT experiments were conducted using slow light at the wavelength (606nm) which is not well adapted for deep tissue imaging [17-18].

In the case of thulium, PSHB can be achieved by applying a static magnetic field, revealing the Zeeman structure [19], as shown in Figure 1(c). As a consequence, atoms relax to their ground state sublevel which acts as a new long lived shelving state. PSHB at convenient operation wavelength is therefore possible using thulium.

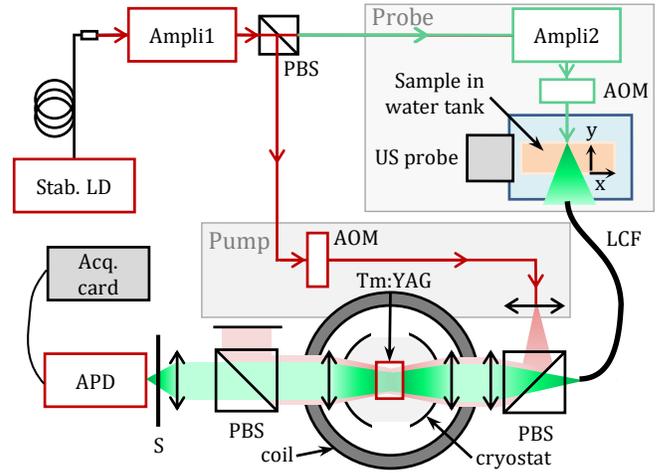

Fig. 2. Optical setup, with a probe beam (green) and a pump beam (red). Stab. LD stabilized laser diode, Ampli 1, 2: laser amplifiers, PBS: polarizing beam splitter, AOM: acousto-optic modulator, US probe: ultrasound probe, x: longitudinal coordinates along US propagation, y: transverse coordinates perpendicular to US propagation, LCF liquid core fiber with high numerical aperture, S: mechanical shutter, APD: avalanche photodiode, Acq. card: acquisition card.

In this article, we report the first demonstration of UOT-imaging based on PSHB in thulium, without slow light. To improve the detection dynamics, two spectral holes are tuned at the two tagged photons' frequencies as illustrated in Figure 1(a). We first compare the SHB and PSHB filters on a weakly scattering sample, and then demonstrate mm imaging resolution on a 1cm scattering sample with two embedded absorbing inclusions.

The experiments are conducted using a 10x10x2 $mm^3$ 2% at. $Tm^{3+}$:YAG single crystal (Scientific Materials Corp.) with an optical thickness close to $\alpha L = 6$ along the 2mm dimension, where $\alpha$ and L are respectively the absorption coefficient and the crystal length. The temperature of the crystal is set at about 2K with a variable liquid helium cryostat (SMC-TBT) so that the homogeneous linewidth is a few kHz, and the measured inhomogeneous bandwidth 30GHz. The 234G magnetic field is created by a pair of water-cooled Helmholtz coils. It is oriented along the [001] crystallographic axis [19]. Under these conditions, the lifetime of Zeeman sublevels is about 30s [20]. Indeed, both heteronuclear cross-relaxation with neighboring Al ions observed at 30 and 60G [21] and spin-lattice relaxation due to phonons observed above 500G [22] are prevented. By controlling the power, duration and spectral width of the pump pulse, the spectral holes width (FWHM) is adjusted to 1.7MHz. This corresponds to a tradeoff between the US pulse width (2.4MHz) and an efficient blocking of untagged photons.

The master laser is an extended cavity diode laser stabilized on a Fabry-Perot cavity [23], leading to a measured frequency drift of about 0.1MHz over 30s. The laser is amplified and split into two optical paths (see Figure 2). In the first path, a 200mW probe beam illuminates the scattering sample, and in the second path a 110mW/$cm^2$ pump beam burns the spectral hole. Both beams are temporally shaped with acousto-optic modulators (AOM, AA Opto-Electronic MT80). US pulses are emitted with a spherical single-element transducer (Olympus 5MHz) of 5cm focal length. The axial resolution calculated for a two-cycle US pulse at 5MHz is 0.6mm and the lateral resolution set by the US waist is 0.6mm.

We prepare two agar gel phantoms of 7x7x1cm³ gels with different intralipid solution concentrations, leading to different scattering coefficients $\mu'_S \approx 1 cm^{-1}$ (sample 1) and $\mu'_S \approx 10 cm^{-1}$ (sample 2), the latter being equal to that of biological tissues [24]. A 2x2x2mm³ black Indian ink inclusion is buried inside sample 1, while sample 2 has two inclusions of 0.5x0.5x0.5mm³ separated by 2mm, close to the sample input surface.

The probe scattered light is collected at the output of the sample with a 1m-long liquid core fiber (LOT-Oriel) with a 45sr.mm² *etendue*. The fiber output is slightly focused into the crystal to illuminate its whole volume. Light exiting the cryostat is collected and focused on an avalanche photodiode (Hamamatsu APD module C12703-01) connected to an acquisition card (Tiepie Handyscope HS5). The *etendue* of the overall collection is limited to 12sr.mm² by the cryostat windows. In addition, the probe light undergoes significant losses between the fiber output and the APD, due to the lenses and cryostat optical apertures (estimated to 70%) and to uncoated cryostat windows (61%).

We choose a co-propagating geometry between the probe and pump beams to maximize their overlap on the crystal. The pump beam is expanded with an 8mm focal lens to approximately match the light cone of the divergent scattered light coming from the liquid core fiber. The two beams are combined with a polarizing beamsplitter (PBS), despite the 50% losses on the probe depolarized light. At the output of the cryostat, another PBS deviates most of the pump power, but the remaining partially depolarized pump light is still transmitted and saturates the APD, therefore it must be blocked with a mechanical shutter (Uniblitz VS14). The AOMs, the ultrasound probe, the mechanical shutter and the acquisition card are all synchronized with an arbitrary waveform generator (Tektronix AWG5004).

Our 2D UOT-images are acquired in $n_l = 51$ lines by translating the US probe in the transverse direction $y$. Each line represents the conversion of temporal traces of the probe pulse into x coordinates, using the ballistic propagation of the US pulse at 1500 m.s$^{-1}$, and it is averaged $n_{avg}$ times. To begin the acquisition of one line, the US probe is mechanically translated in about $T_{mech} \approx 185$ms to its position. One acquisition cycle is depicted on Figure 3(a) and defined as follows: the shutter remains closed while the pump pulse burns 2 spectral holes in the Tm:YAG crystal at ±5MHz from the carrier frequency during 10ms (5ms per hole). The shutter then opens in 4ms, and a series of US pulses are emitted every $T_{US} = 0.1$ms while the sample is continuously illuminated by the probe beam. The shutter closes at the end of the cycle. The key parameter to quantify the imaging efficiency is $n_p$, the number of US pulses emitted per cycle. In the case of SHB, the 8ms lifetime allows only $n_p = 5$ pulses per cycle, above which the exponential decay of the hole transmission degrades the signal. In the case of PSHB, the hole lifetime is 3 orders of magnitude longer, allowing for a commensurate increase of $n_p$, e.g. $n_p = n_{avg}$ for the following images. Figure 3(b) shows the decay of the recorded tagged photons when the pump beam is blocked at $t = 0$. After a rapid 25% decay due to partial storage in the metastable level, the filter transmission loses another 20% in 10s, which is consistent with the 35s spectral hole lifetime measured using independent time-resolved spectroscopy. Tagged photons are attenuated by a factor of -20dB, which is 6dB less than untagged light. One cycle lasts $T_{cy} = 100ms$ as set by the shutter repetition rate, allowing a faster frame rate than in the case of a cycle with a period equal to the hole lifetime. At the end of each cycle, data are transferred to the computer. One cycle will be repeated $n_{cy} = n_{avg}/n_p$ times to reach $n_{avg}$ averaging for each line.

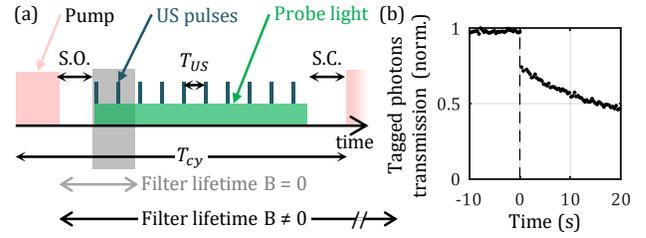

Fig. 3. (a) Pulse cycle for UOT. The pump pulse (red) creates the 2 spectral holes at the tagged photons frequency. The simultaneous ultrasound pulse (blue) and probe light (green) create the tagged photons at $T_{US}$ repetition period. When $B = 0$, only the 5 first US pulses are exploitable due to the short filter lifetime (gray area). SO: shutter opening, SC: shutter closing. (b) Transmission of the normalized tagged photons.

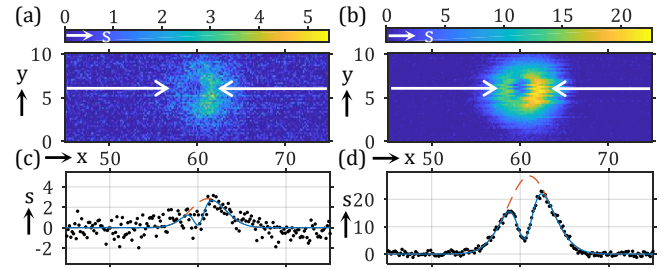

Fig. 4. UOT image of sample 1 ($\mu'_S = 1\ cm^{-1}$) (a) with a spectral hole (no magnetic field) and (b) with a persistent hole (234G magnetic field). Each image is averaged 100 times. (c) and (d) are the profiles along the white arrows in images (a) and (b) respectively. Black dots: data, blue line: fit, red dashed line: Gaussian fit (see text for details), x: longitudinal coordinates (mm), y: transverse coordinates (mm) and s: transmitted light detected by the APD (mV).

We first image sample 1 to study the effect of the magnetic field on the imaging performance. SHB and PSHB images are respectively shown on Figure 4(a) and (b), with each line averaged $n_{avg}$ =100 times. The first striking result is that $T_{acq}^{SHB} = 8.5$min are required to generate an image using SHB while this time is reduced to $T_{acq}^{PSHB} = 30$s in the case of PSHB. Since data are in the present case digitized after each cycle, which is time consuming especially for high $n_{cy}$, for a fair comparison we calculate the relevant time dedicated to imaging only. It fundamentally reduces to $T_{img} = (T_{mech} + n_{cy}T_{cy})n_l$ and gives a duration of $T_{img}^{SHB} = 111s$ ($n_{cy} = 20$) and $T_{img}^{PSHB} = 14s$ ($n_{cy} = 1$). Hence the number $n_p$ of successive US pulses which can be emitted without refreshing the hole is crucial for fast acquisition. The delay $T_{mech}$ remains nonetheless a limiting time and could be reduced using a transducer array to perform electronic rather that mechanical scanning in the transverse direction, e.g. one US pulse every $T_{US} = 0.5$ms and $T_{mech} = 0$. The acquisition time would then be shorten by $T_{mech} \cdot n_l = 185$ms $\cdot 51 = 9$s.

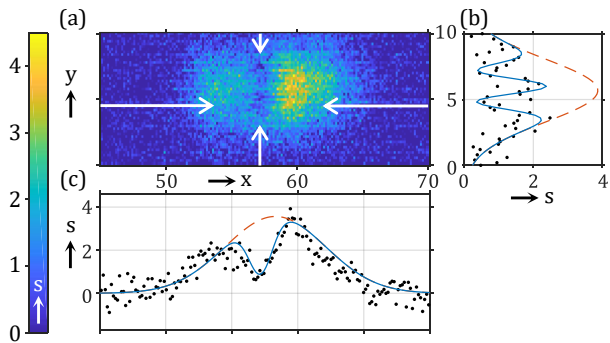

Fig. 5. Imaging resolution with persistent SHB (a) Image of sample 2, averaged 200 times. (b) Profile along $x = 57mm$. (c) Profile along $y = 4.4mm$. Black dots: data, blue line: fit, red dashed line: Gaussian fit (see text for details), x: longitudinal coordinates (mm), y: transverse coordinates (mm) and s: transmitted light detected by the APD (mV).

The second striking result is that the contrast-to-noise ratio (CNR) of the inclusion increases from 3.3±0.6 for SHB to 32±0.7 for PSHB. This is due to a tenfold improvement of the spectral holes transmission. Therefore, with this experiment, we demonstrate the double benefit of PSHB filter in terms of hole depth (resulting in an increase of the tagged photon signal) and imaging speed.

In a second experiment, we evaluate the spatial imaging resolution of our PSHB-UOT setup. In Figure 5, we show a UOT image of sample 2, averaged over 200 single images corresponding to $T_{acq}^{PSHB} = 47s$ acquisition time. The inclusions are clearly visible despite a modest CNR ($7.5 \pm 0.4$). This CNR is mainly limited by the strong scattering coefficient of sample 2 ($\mu_s'$=10cm$^{-1}$), leading to a lower signal collection compared to sample 1. The image is fitted with a wide 2D-Gaussian function representing the beam envelope minus a pair of narrower 2D-Gaussian functions for the inclusions (same method used for Fig. 4). We obtain a FWHM size (1.8±0.2mm)$^2$, significantly larger than the (0,5mm)$^2$ initial inclusions size at the time of fabrication. This is essentially explained by the finite US resolution (0.6mm) and by the possible uncontrolled diffusion of the ink inside the gel due to chemical similarity of the inclusions and the surrounding gel. In spite of a modest CNR, this experiment validates the possibility of using Tm:YAG PSHB filter to image millimeter-sized inclusions in strongly scattering samples.

The performance of this proof of principle is mostly limited by technical issues. Indeed, in addition to switching to an ultrasound probe array to reduce the scanning time, one can improve the CNR by increasing the light probe power within the medical safety limit [25] and anti-reflection-coated cryostat windows and crystal. Furthermore, a larger pump power and a more absorbing Tm:YAG crystal would lead to a deeper spectral hole, and a larger CNR. Finally, although the *etendue* of our setup is close to the state of the art [17], it could still be improved by using a cryostat with better optical access.

In summary, we have conducted the first experimental demonstration of UOT imaging using a filter based on PSHB in Tm:YAG, by applying a magnetic field on the crystal maintained at 2K. Our experiment performed on scattering phantom gels shows the possibility to perform *in vivo* cm-depth imaging of scattering tissues at 793nm. Indeed, compared to anterior demonstrations deeper spectral holes increases the number of detected tagged photons by a factor 10, leading to a larger CNR. In addition, the extended hole lifetime allows up to 40 times more acquisitions per refreshing cycle. Our work therefore opens the possibility of UOT imaging at video rate. Such a probe will be integrated in our future UOT setup where Tm:YAG PSHB detection will be tested on actual living tissues.

**Funding.** ITMO Cancer AVIESAN MALT project (C16027HS); CNRS (Défi Instrumentation aux Limites 2016); LABEX WIFI (ANR-10-LABX-24 and ANR-10-IDEX-0001-02 PSL*).